\documentclass[12pt, draftclsnofoot,onecolumn]{IEEEtran}
\usepackage{graphicx}
\usepackage{subfigure}
\usepackage{cite}
\usepackage{amssymb, amsmath}
\usepackage{multirow}
\usepackage{xcolor}

\begin{document}

\title{Bi-Directional Mission Offloading for Agile Space-Air-Ground Integrated Networks}

\author{Sheng~Zhou, Guangchao Wang, Shan~Zhang, Zhisheng~Niu, Xuemin~(Sherman)~Shen
\thanks{Sheng~Zhou, Guangchao~Wang and Zhisheng~Niu are with Beijing National Research Center for Information Science and Technology, Department of Electronic Engineering, Tsinghua University, Beijing 100084, China.

Shan~Zhang (\emph{corresponding author}) is with the School of Computer Science and Engineering, Beihang Univesity, Beijing 100191, China. She is also with the State Key Laboratory of Software Development Environment, and the Beijing Key Laboratory of Computer Networks.

Xuemin~(Sherman)~Shen is with the Department of Electrical and Computer Engineering, University of Waterloo, 200 University Avenue West, Waterloo N2L 3G1, Ontario, Canada.
 }}

\maketitle

\begin{abstract}
Space-air-ground integrated networks (SAGIN) provide great strengths in extending the capability of ground wireless networks. On the other hand, with rich spectrum and computing resources, the ground networks can also assist space-air networks to accomplish resource-intensive or power-hungry missions, enhancing the capability and sustainability of the space-air networks. Therefore, bi-directional mission offloading can make full use of the advantages of SAGIN and benefits both space-air and ground networks. In this article, we identify the key role of network reconfiguration in coordinating heterogeneous resources in SAGIN, and study how network function virtualization (NFV) and service function chaining (SFC) enable agile mission offloading. A case study validates the performance gain brought by bi-directional mission offloading. Future research issues are outlooked as the bi-directional mission offloading framework opens a new trail in releasing the full potentials of SAGIN.
\end{abstract}

\section{Introduction}
By interworking the communication and computing resources on satellites, high-altitude platforms (HAPs), unmanned aerial vehicles (UAVs) and terrestrial wireless communication nodes, the space-air-ground integrated networks (SAGIN) is expected to exploit the advantages of each component network, as well as provide wide-range and seamless networking services. Many applications, that require real-time and reliable data sensing, collection, transmission, processing and distribution across large areas, can be supported, such as earth observation, navigation, smart cities, connected vehicles, and global wireless network coverage.
Recent studies have revealed the great potentials of using satellite networks to assist ground wireless networks for ubiquitous access \cite{Sat5G}, and to handle communication missions that can hardly be accomplished solely by the ground network \cite{Bai5G}, e.g., efficient broadcasting and multicasting for connected vehicles \cite{ZhangN17}. This mission offloading from ground to space makes use of the wide coverage and reduces the communication delay through fewer hops.

However, the resource in the space-air networks is often scarce. The spectrum is limited, and the processing capability is constrained by the computing resources and the power supply on the satellites, HAPs and UAVs. Under many circumstances, stand-alone space-air network can barely accommodate emerging missions calling for a large amount of data acquisition and dissemination, and intense computing tasks requiring real-time processing \cite{Wu16}. To this end, collaboration among multiple satellites from different orbits has been proposed to achieve better resource utilization and accordingly improved performance \cite{Du16}. In fact, the ground network can also help accomplish the missions in the space-air networks, exploiting its rich spectrum, computing and storage resources, i.e., missions originally carried out in space-air networks can be \emph{reversely offloaded} from the space-air to the ground. For instance, computing intensive tasks, such as remote sensing and cooperative monitoring, can be offloaded from space-air nodes to servers on the ground. In coordination with the existing mission offloading from the ground to the space, this bi-directional mission offloading in SAGIN can make full use of the complementary advantages of both space-air and ground networks.

However, the success of bi-directional mission offloading in SAGIN highly relies on abilities to coordinate the heterogeneous resources from satellites, HAPs, UAVs and ground networks, which is far more challenging than managing resources within individual networks. In addition, to adapt itself to the highly dynamic environment, the integrated network needs technologies like software defined network (SDN), network function virtualization (NFV), cross-layer resource management and network reconfiguration. Recently, SDN has been proposed to realize flexible resource management and resource-mission matching, for earth observation missions \cite{Sheng17} and space-assisted connected vehicles \cite{ZhangN17}. Due to the large network scale of SAGIN, realizing SDN must face the challenges of system information collection. While for agile offloading on both directions in SAGIN, not only the network functions (NFs) have to be virtualized and matched to the heterogeneous resources, but also the service function chain (SFC) must be planned and optimized with the network management. As a result, a unified framework supporting SAGIN reconfiguration is required.

In this article, we first introduce the concept of bi-directional mission offloading and its potential applications. Following the review of the NFV and SFC technologies used in wireless networks, we illustrate how they can be implemented in SAGIN to enable the agile bi-directional mission offloading and network reconfiguration. We then elaborate our idea with a case study that addresses SFC and resource provisioning in SAGIN. We conclude the article with outlook on promising research issues.

\section{Bi-Directional Mission Offloading: Making the Best Use of a 3D Network}

    \begin{figure}[!t]
			\centering
			\includegraphics[width=6.5in]{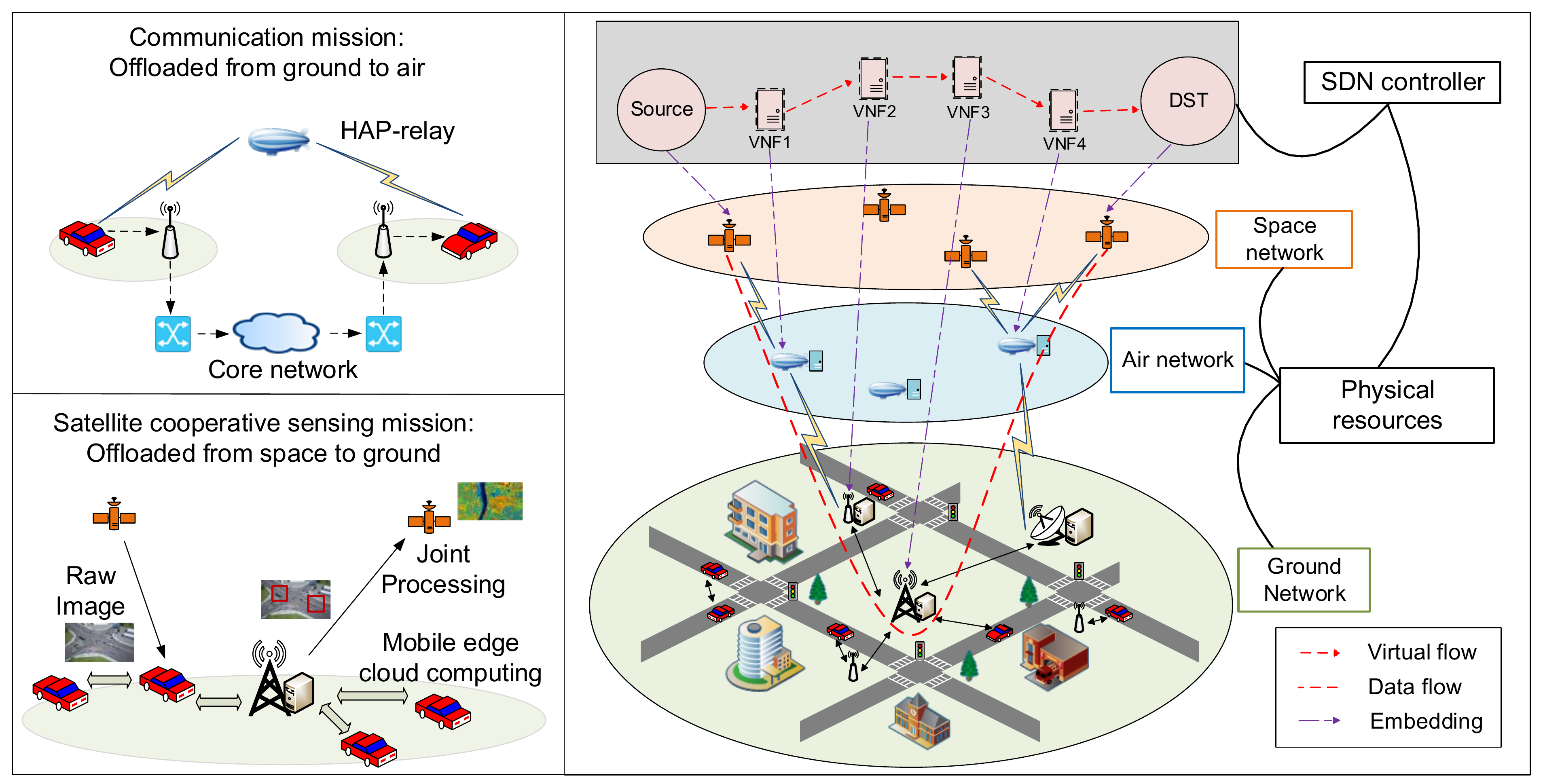}
			\caption{Use cases of bi-directional mission offloading in SAGIN, with examples of communication missions offloaded from ground to air and computation mission offloaded from space to ground, respectively. The general concept of VNF embedding and SFC planning for agile offloading is illustrated, with mission offloading from space to ground.}
			\label{SFC_en}
	\end{figure}

	The SAGIN is expected to exploit the complementary advantages of space, air and ground facilities to provide reliable, high data rate, and cost-efficient services with seamless coverage. An example is illustrated in Fig.~\ref{SFC_en}.
	Specifically, the facility-rich and low-cost ground network will provide services for the majority population in urban areas, while the large-coverage space-air facilities can help uncovered areas or special events \cite{ZhangN17}.
	Based on the distinct features of different facilities, closer collaborations are required to make the best use of this 3-dimensional network.
	To this end, we propose to conduct agile bi-directional computation and communication mission offloading, which can benefit the performance of \emph{both} ground and space-air networks. Specifically, ground missions correspond to missions (mainly communications) originally carried out in ground networks, while the space-air missions correspond to the missions (mainly computation and processing) originally carried out in space-air networks. The label of the missions (ground or space-air) does not change even if they are offloaded in SAGIN.

	\subsection{Ground Missions Offloaded to Space-Air}
       Flying at altitudes of 17$\sim$22 km, HAPs can provide coverage of hundreds of kilometers with low path-loss, and has been granted with 600 MHz dedicated bandwidth at 47--48 GHz band for communications by the Federal Communications Commission (FCC).
       Compared with HAPs, UAVs also enjoy better mobility control in addition to Line-of-Sight (LoS) coverage.
       The battery life, typically 30 minutes to 6 hours, is the main technical concern of UAVs, while the energy harvesting technologies can relieve this pressure\footnote{https://www.theguardian.com/technology/2015/jul/31/facebook-finishes-aquila-solar-powered-internet-drone-with-span-of-a-boeing-737}.
       The broadband communication satellites (running at low, medium, or geostationary orbits) can provide globally seamless coverage, and the capacity is also improving with technologies such as multiple beam antennas, advanced onboard processing and control methods.
       In addition, the rapid development of space-air technologies greatly reduces the costs of HAPs, UAVs, and satellite communications, and can potentially bring space-air network access into daily life.
       The Loon Project aims to provide Internet access through HAPs with comparable speed of 4G and long term evolution (LTE) networks, which has been implemented for the after-disaster reconstructions in 2017.
       The Globe Xpress system employs three geostationary-orbit satellites to provide broadband network access for 99\% global coverage, with downlink and uplink rates of 50 Mbps and 5 Mbps, respectively.

       The ground network, with well-developed technologies and infrastructures, will take the main responsibility to provide cost-effective access in general.
       Meanwhile, the ground communication missions can still be offloaded to space-air networks on-demand, bringing numerous benefits.

       \textbf{1. Coverage and Capacity Enhancement.} With the large coverage and dedicated bandwidth, HAPs are promising to construct low-cost rural area coverage. The satellites can further cover mountains and seas to form the global umbrella coverages. In addition to coverage enhancement, the air and space networks can also improve network capacity through large-area broadcast, such as location-based services and vehicle software update.

       \textbf{2. Mobility Supports.} The emerging vehicular communication has posed great challenges for ground networks on high mobility, when supporting the latency-critical road-safety applications. The large-coverage air and space networks can reduce the handover frequency, enhancing the service reliability. Furthermore, the UAVs can even form moving cells to vehicles, providing handover-free network access.

       \textbf{3. Robustness.} The ground networks are vulnerable to disasters, and the recovery is also time-consuming. In comparison, HAPs and UAVs can be rapidly deployed on-demand, while the coverage of satellites is robust to disasters.

       \textbf{4. Wireless Backhauling.} As network densifies, the backhaul will become a key bottleneck. The space-air networks can provide high-speed backhaul access for ground base stations (BSs) through mmWave communications, liberating the wired backhauls. Accordingly, ground stations can be deployed in a flexible plug-and-play manner.

	\subsection{Space-Air Missions Offloaded to Ground}

	Driven by the need for supporting diversified missions, the space-air facilities will be implemented with on-board processing units for functions like intelligent sensing and information processing.
	However, the on-board processing units can bring more energy consumptions and the increased weights.
	The ground network, especially considering the emerging autonomous-driving vehicles, owns abundant computation and storage resources.
	Accordingly, the inverse directional mission offloading enables space-air facilities to make use of ground resources, and promotes the space-air intelligence and sustainability. One such offloading example is shown in Fig.~\ref{SFC_en}. Two satellites are employed to jointly monitor and detect abnormal ground conditions, where one satellite offloads the computation intensive image processing missions to the ground network. Through mobile edge computing, the BSs and vehicles analyze the raw images cooperatively by utilizing their available general-purpose computation resources. The processing results are fed back to the other satellite to make further analysis and actions.
	With recent developments of NFV and SDN, space-air NFs are decoupled from the hardware and thus flexible mission offloading can be made possible, bringing following advantages.	
	
	\textbf{1. Enhanced sustainability.}
	As the computation-intensive on-board processing can cause high power consumption, the computation mission offloading can effectively reduce the power consumption of space-air facilities to prolong their battery life.
	In addition, for the renewable energy powered space-air facilities, the computing missions can be dynamically offloaded to ground based on the available battery, energy harvesting rate and on-board processing load, so as to enhance the system reliability and sustainability.
	
	\textbf{2. Improved intelligence.}
	Offlading computation-intensive missions from space-air to ground nodes, either fixed BSs or moving vehicles, for timely processing, has the similar effect as enhancing the on-board processing capability of space-air facilities.
	The enhanced processing capability provides new opportunities to develop space-air intelligence, such as autonomous flying, monitoring/observatoin, localization and tacking.
	Furthermore, the data collected from space-air and ground nodes can be processed jointly to enhance inference and detection precision, for example in the accurate positioning of UAVs with high-definition map.
	
	\textbf{3. Simplified facility design.}
	As missions can be offloaded to the ground, the simplified hardware design is possible for space-air facilities.
	Accordingly, the weights of space-air nodes can be significantly reduced, with enhanced battery life and thus flying sustainability.
	In addition, the costs of nodes can be reduced, making it feasible to construct large-scale yet cost-effective space-air networks.
	Similar idea has been implemented on ground networks, i.e., the architecture of cloud radio access network (CRAN), where the remote radio head only keeps radio frequency transmission functions while the baseband processing is conducted in data centers.

\section{Reconfiguration via NFV and SFC: Enabling Agile Mission offloading}
	The fundamental issue of bi-directional mission offloading in SAGIN is how to handle diverse missions with heterogeneous resources from space, air, and ground.
	Also, SAGIN shows high dynamics due to the flexible deployment and mobility of not only air and space nodes, but also ground nodes like vehicles.
	Thus, the SAGIN should be capable of reconfiguration along with the variations of network status and traffic demands. Flexible reconfiguration requires decoupling NFs from hardware, which can be accomplished by NFV.
    Moreover, due to the large scale of SAGIN, it is hard to ensure the coexistence of all kinds of resources in need. As a result, the NF for certain mission cannot be supported anywhere in SAGIN, for instance, observation functions are not supported by ground nodes, while the storage and computing resources may not be sufficient on the space-air nodes. This actually calls for the implementation of SFC on top of NFV, that geographically matches the resource to virtual network functions (VNFs). The signalling overhead for collecting necessary system information for SFC is then vital, and hierarchical control \cite{ZhangN17}, both in space-air and ground networks is an option to alleviate the overhead.
	Albeit born with wireline network, the NFV and SFC techniques enable such agility, whereby the NFs can be orchestrated on demand and embedded based on the resource availability, as shown in Fig.~\ref{SFC_en} for the offloaded mission from the space-air network to the ground network.

	\subsection{NFV and SFC in Wireless Networks}
		NFV can decouple the network services from the dedicated wireless network hardware, the standardization and implementation of which are provided by the ETSI Industry Specification Group \cite{ETSI13}. Using NFV, the traditional NFs are virtualized into software components {as VNFs} that run on universal hardware. The physical wireless network resources, such as radio spectrum, the CPU cores, and network infrastructure, are abstracted as virtual wireless network resources and allocated to various VNFs to hold various network services \cite{Ye18}. In this way, the VNFs can be flexibly deployed and reconfigured on different physical nodes which facilitates the network agility and robustness. In addition, better network scalability is achieved because new VNFs can be easily added to support new services. As the VNFs can be shared among multiple service requests, both communication resources and computation resources are utilized effectively. In current IP-based LTE architecture, the NFV can be applied to multiple network segments including radio access networks (RANs), evolved packet core (EPC) networks and transport networks \cite{Liang15}. In general, following NFs can be virtualized by NFV:
		
		\begin{itemize}
			\item RAN functions such as baseband processing functions
			\item EPC functions such as public data network gateway (P-GW), serving gateway (S-GW), mobility management entity (MME)
			\item Security related functions such as Firewall, Intrusion Detection Systems (IDS), Deep Packet Inspection (DPI)
			\item Network optimization functions such as Load balancer, traffic control protocol optimizer, Network Address Translator (NAT)
		\end{itemize}
		
		With the advances in NFV technology, the concept of SFC is defined by Internet Engineering Task Force (IETF) \cite{IETF15}, which also demonstrates the SFC applications in mobile networks. In this paradigm, the data traffic is required to flow through several specific NFs under a specified order for proving the network services with heterogeneous quality-of-service (QoS), security and reliability requirements \cite{WNV-Yu15}. Conventionally, NFs are implemented using dedicated hardware. This coupling of functions and hardware leads to a static SFC, which typically can not share the functions with other network services. Thanks to the decoupling of the functions and hardware, NFV technology enables the agile SFC and resource sharing by virtual functions.

	\subsection{Reconfiguration in Space-Air-Ground Integrated Networks via NFV and SFC}

		\begin{figure}[!t]
			\centering
			\includegraphics[width=5.6in]{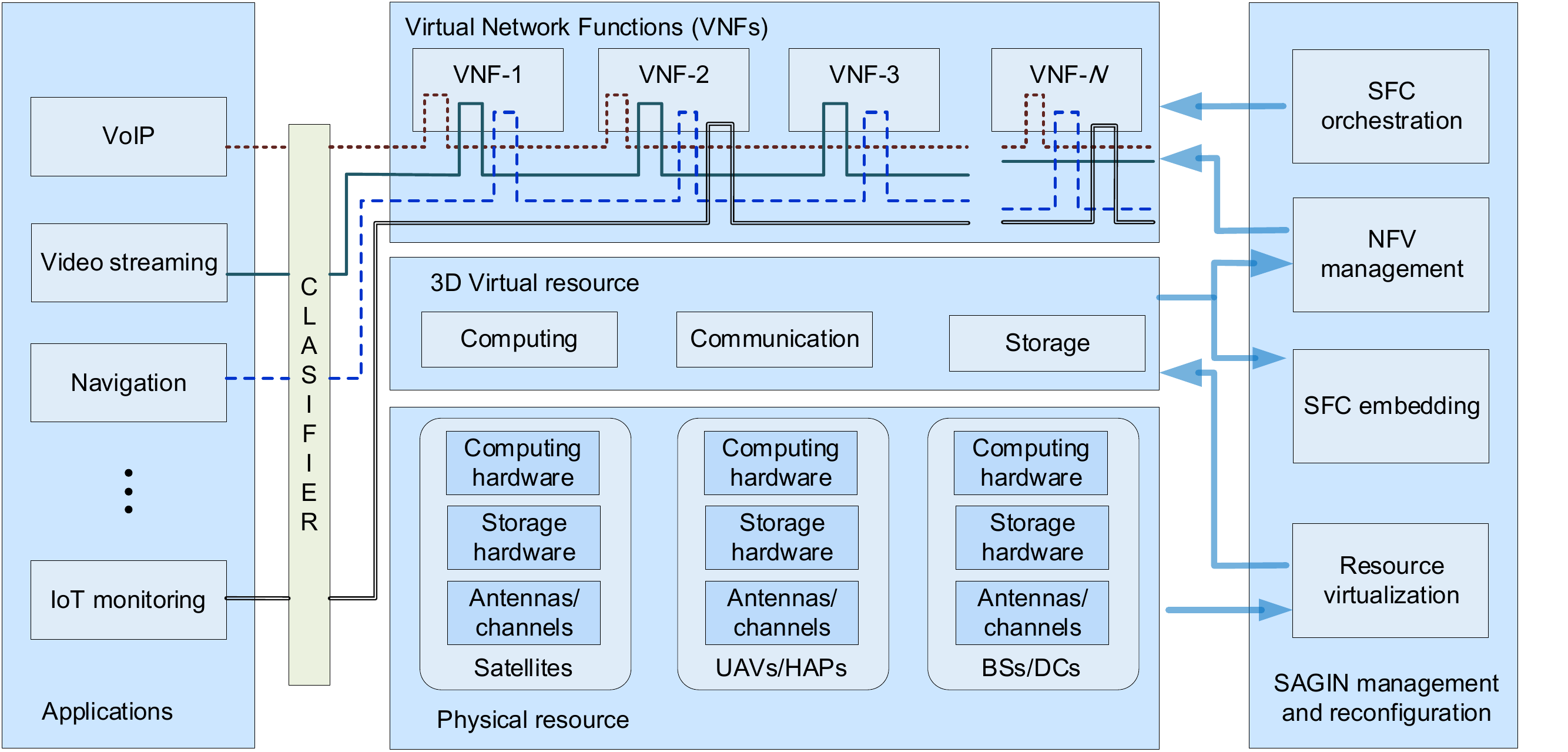}
			\caption{SAGIN management and reconfiguration for bi-directional mission offloading.}
			\label{control_block}
		\end{figure}

		By decoupling NFs from hardware, NFV and SFC enable agile network reconfiguration to deal with network and traffic dynamics.
		The framework for network reconfiguration is shown as Fig.~\ref{control_block}.
		Each SAGIN application can be classified based on the QoS demands, whereby the required NFs are orchestrated as an SFC to support the application.
		The SFC embedding is used to map the NFs of each chain to the SAGIN physical infrastructures through NFV management, which is the main challenge in implementation \cite{Xie16}. NFV enables the flexible placement of VNFs on different physical nodes, which brings dynamic resource allocation problems. The installation and migration problems of VNFs are often NP-hard, hence are non-trivial to solve for large-scale instances. So it is imperative to exploit efficient approaches to obtain optimal dynamic resource allocation.
		Thus, the resource virtualization function is introduced, whereby the heterogeneous physical resources (e.g., power, spectrum, computation, storage) from SAGIN are abstracted as virtual resources (e.g., transmission rate, computation frequency, cache size).
		With resource virtualization, the space, air, and ground networks can be seen as unified components, which helps to ease the SFC embedding.
		
		The SAGIN demonstrates high dynamics in three aspects: the application and traffic demand variations, the flexible deployment of HAPs and UAVs, the mobility and availability of space-air and ground nodes.
		Such dynamics require SAGIN be reconfigured agilely, through re-optimizing the relationship between SFCs and physical resources based on the network and traffic status.
		As such, the SFCs can be better supported with high reliability and resource efficiency.
		However, the spatial, temporal, and network scales need to be carefully judged to balance SFC embedding performance and reconfiguration costs, mainly caused by system information collection. The entities carrying SAGIN management and reconfiguration on the right of Fig.~\ref{control_block} can have hierarchical relations and be distributed over SAGIN.
		We will illustrate the basic idea of enabling bi-directional mission offloading with SFC-based SAGIN reconfiguraion in the following case study, and then discuss possible research issues in the next section.

	\subsection{A Case Study}

\begin{figure}[!t]
		\centering
		\includegraphics[width=4.0in]{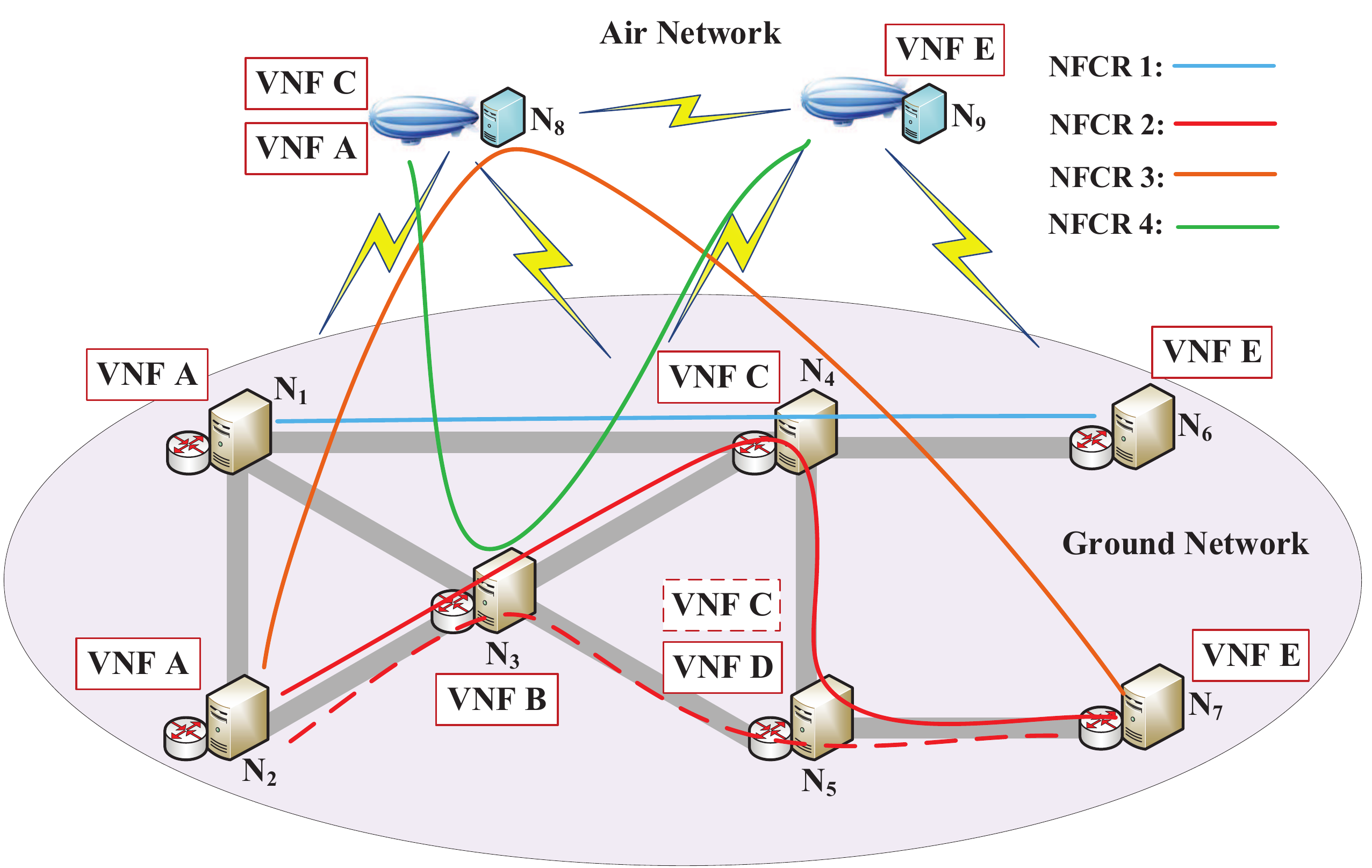}
		\caption{An example of a space-air-ground integrated network topology.}
		\label{SFC_ex}
	\end{figure}

	SFCs are established and deployed for missions that are offloaded in SAGIN. With NFV technology, NFs are implemented by VNFs, which can be flexibly embedded in both air nodes and ground nodes.
	We consider an integrated network with two air nodes and seven ground nodes, as illustrated in Fig. \ref{SFC_ex}. Thanks to the large coverage, the air nodes have full connections with the ground nodes. The figure shows 4 examples of network function chain request (NFCR) which represent resource provision strategies for corresponding missions. NFCR 1 requires VNF A, C and E, where VNF C is embedded on node N$_{4}$. The traffic of NFCR 1 showed in blue line originates from node N$_{1}$ and eventually flows to node N$_{6}$. NFCR 2 requires VNF A, B, C, D and E to complete the mission. The red line shows one chaining strategy of NFCR 2, where the traffic experiences 4-hop transmissions. Thus VNF C is shared between NFCR 1 and NFCR 2, and the computation cost for installation and maintenance of VNF C can be saved. The red dot line shows another strategy, where the traffic only experiences 3-hop transmissions so that the bandwidth cost is reduced but an additional VNF C instance should be embedded in N$_{5}$ which produces additional computation cost. Therefore there exists a tradeoff between the computation cost and the bandwidth cost. The required VNFs of NFCR 3, as a ground network mission, are partially offloaded to air nodes as illustrated in the orange line. The traffic only experiences 2-hop transmissions that effectively reduces the hopping delay and potential traffic congestion can be avoided in a crowded ground network scenario. Conversely, the VNFs with high computation complexity can be offloaded from the air nodes to ground nodes to save the computation resources and relieve the traffic burden, as shown in the green line.
	
	Here the key issue is the bandwidth and computation resource provision for flexible NF chaining. Thus the VNFs mapping and traffic assignment should be jointly considered. The VNFs mapping problem is how to embed the required VNFs to proper physical network nodes. The traffic assignment problem is how to steer the traffic flow through VNFs in a specified order over proper physical links. The major constraint is the resource capacity of the physical network including bandwidth resources and computation resources. The objective is to maximize the number of mission requests that can be successfully served and minimize the bandwidth and computation cost. This problem can be formulated as a nonlinear integer programming problem, which is solved by Matlab SCIP toolbox. In the simulation, we consider delay sensitive and computation intensive missions which are randomly generated with 3-6 VNFs in both ground network and air network. The bandwidth capacity of links and computation capacity of nodes are uniformly distributed in $[80,100]$ Mbps and $[500,600]$ GFLOPS\footnote{GFLOPS: Giga Floating-point operations per second; }, respectively. The delay of each hop is set to $[10,15]$ ms following uniform distribution. Each air node is equipped with a $100$ Wh battery that exclusively supplies power for the computation of the missions. Each VNFs consumes $0.2$ W power for one mission. We assume that the energy is sufficient for the propulsion systems and the \emph{duty endurance time}, defined as the time duration that the energy for processing the missions is sufficient, depending on the endurance of the computation battery.

	\begin{figure}[!t]
		\centering
		\includegraphics[width=4.0in]{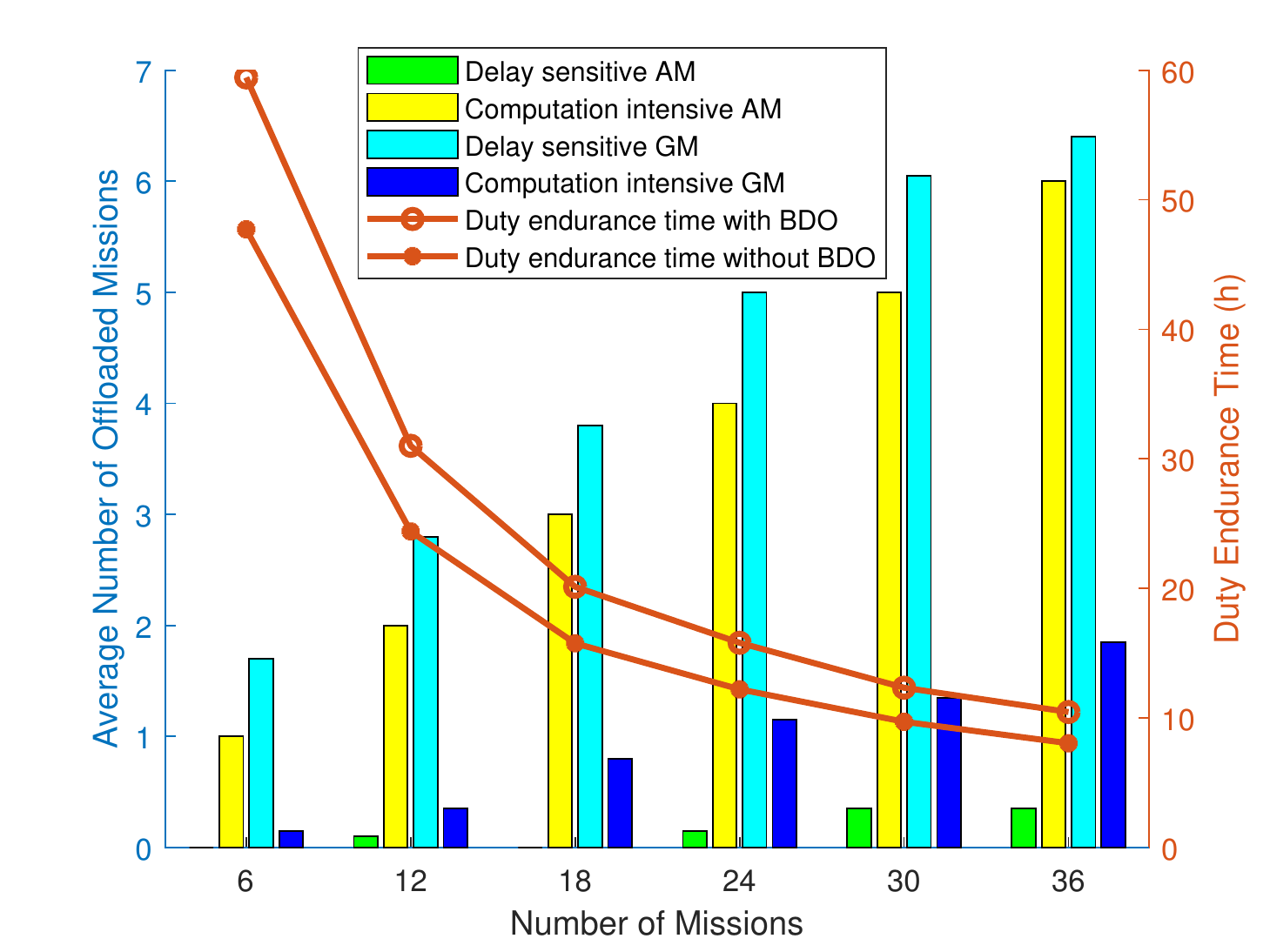}
		\caption{Performance of bi-directional offloading (BDO) for air mission (AM) and ground mission (GM).}
		\label{sim_off}
	\end{figure}
	
	\begin{figure}[!t]
		\centering
		\includegraphics[width=4.0in]{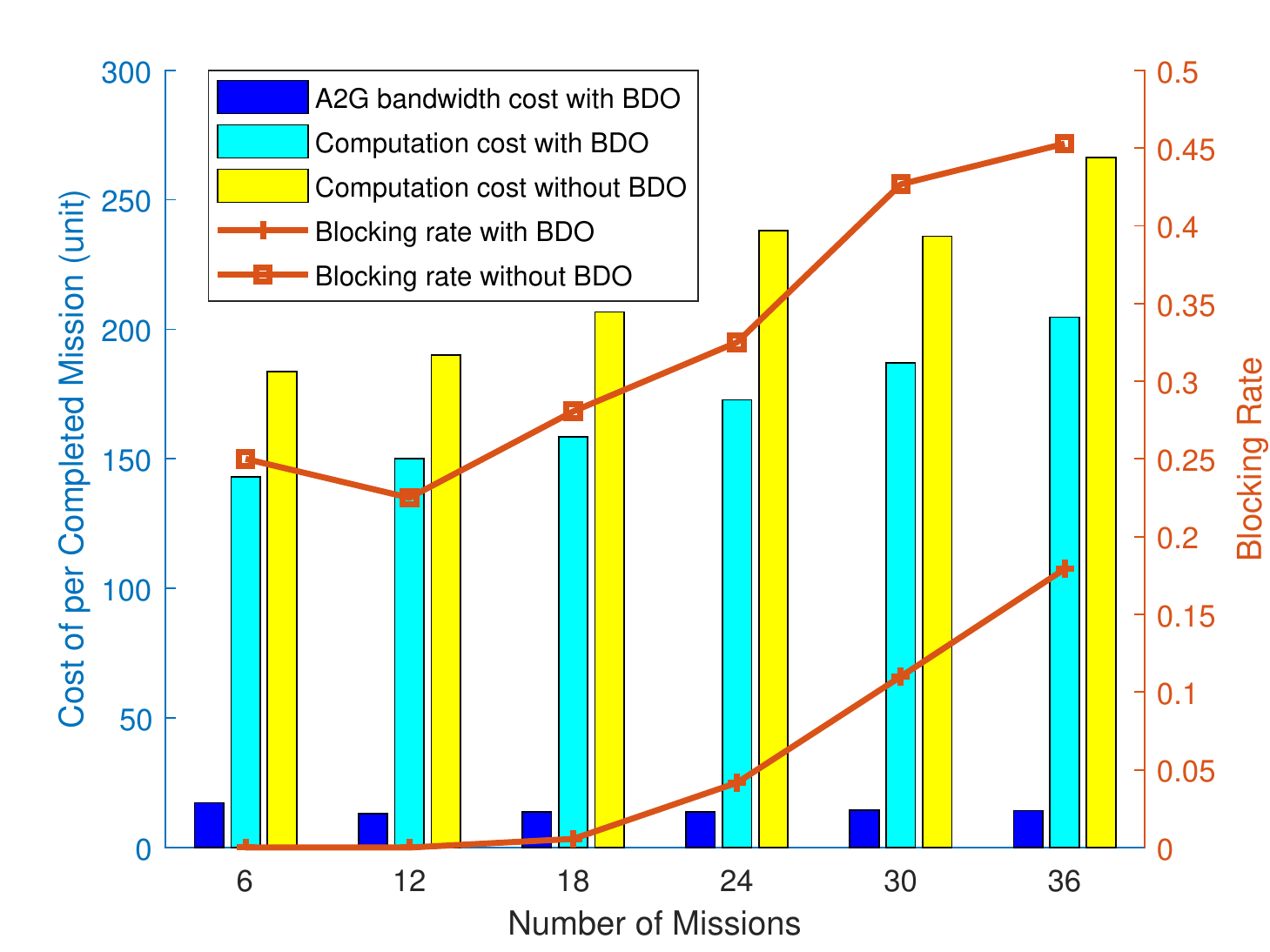}
		\caption{Resource cost and blocking rate with or without bi-directional mission offloading.}
		\label{sim_com}
	\end{figure}
		
	The performance of bi-directional mission offloading is shown in Fig. \ref{sim_off}. The delay sensitive missions have more opportunities to be offloaded from the ground nodes. This is mainly because the offloaded missions only experience 2-hop transmissions and the hopping delay can be reduced. However, among the missions from the air network, the computation intensive missions are preferred to be offloaded due to the fact that the computation cost is lower in ground nodes and more computation resources in the air can be saved. The duty endurance time of air nodes decreases with the increase of the number of missions, while the duty endurance time of the air nodes notably increases by bi-directional mission offloading. This is mainly because of reversing mission offloading from air to ground, which leads to the reduction of energy consumption for computation.
	
	Fig. \ref{sim_com} compares the performance with and without bi-directional mission offloading. The figure shows that the computation resource cost per completed mission is substantially reduced by bi-directional mission offloading, indicating more efficient resource utilization of the whole network. However, additional bandwidth cost for air to ground (A2G) links is introduced. Furthermore, the network with bi-directional mission offloading has a lower blocking rate, as a result of more agile resource management enabled by SAGIN network reconfiguration.
	
	In summary, the case study validates the feasibility of implementing bi-directional mission offloading in SAGIN via NFV and SFC, through formulating and solving the VNFs mapping and traffic assignment problems. Simulations confirm that the bi-directional mission offloading significantly improves the resource utilization efficiency and the sustainability of air nodes.

\section{Research Issues of Bi-Directional Mission Offloading}
In this section, we highlight key research issues to be solved before realizing the concept of SFC-based SAGIN reconfiguration and supporting bi-directional mission offloading.

\subsection{Abstraction and Virtualization of Heterogeneous Resources}
In SAGIN, the resources include the wireless spectrum, computing processor, storage, and observation resources, and etc. Therefore, it is vital to virtualize these heterogeneous network resources as the basis of SDN and NFV, in order to enable bi-directional offloading. These heterogeneous resources should have a unified abstraction which satisfies the needs for network virtualization, including flexibility, isolation and coexistence \cite{WNV-Yu15}. Flexibility highly relies on how resources are abstracted. One way is to use the information bits that can be processed, i.e., transmitted, computed, stored and observed, via various network resources \cite{Sheng17}. This approach is mission dependant, and may not be easily generalized to different missions. For example, tasks can have different computing complexities even if their input bits are of the same amount. In addition, delay performance can hardly be reflected. Here, possible approach is to sacrifice certain flexibility for better resource virtualization precision. Taking delay as the major performance metric, the effective bandwidth can be used to abstract the realtime transmission capability of wireless resources, and the computation frequency like FLOPS and DMIPS\footnote{DMIPS: Dhrystone million instructions executed per second.} for computing resources with computing delay as the computation complexity divided by the frequency. As for isolation requirements, computation, storage and wireless resources in space and air are mostly easy to be isolated and sliced. Wireless resources on the ground require more research efforts \cite{NVS12}. While the resources in large-scale SAGIN do not in general coexist, jointly optimizing SFC and mission offloading can potentially act as if the resources are available everywhere.

\subsection{Service Function Chaining and Placement}
When the accomplishment of a certain mission spreads over different geographical locations, for instance, the start and the end points of the mission are not at the same spot, or the resources needed to support the mission are not co-located, SFC can plan the component VNFs of the mission and their corresponding virtualized network resources. The placement of VNFs is subject to the geographical distribution of computing, storage and observation resources, as well as the bandwidth resources that connect them. The resources may also be shared and sliced by multiple VNFs from different missions. End-to-end processing delay of the mission serves as the performance constraint. The corresponding optimization method has been illustrated in our case study. In addition, one key challenge is to exchange the resource information among space, air and ground networks, this can lead to high signaling overhead. Reducing such overhead can cause imperfect system information, or prolong the acquisition delay, both of which can degrade the performance of SFC and placement.

\subsection{Reconfiguration and Resource Scheduling}

	\begin{table}[!t]
		\caption{Heterogeneous resource management for SAGIN reconfiguration }
		\label{reconfig_table}
		\centering
		\begin{tabular}{|c|c|c|c|}
			\hline
			\hline
			\textbf{Resource} & \textbf{Reconfiguration Trigger} & \textbf{Time Scale} & \textbf{Scheduling Basis}\\
			\hline
            \hline
			Time-frequency radio & Channel status variations &  Seconds to minutes & Interference-aware\\
			\hline
			Pilots & User mobility &  Seconds to minutes & N/A\\
			\hline
			Backhaul bandwidth & Baseband function splitting & Minutes & Co-design with caching\\
			\hline
			Computation & New task arrivals & Minutes & Computation task offloading \\
			\hline
			Storage & Content popularity and requests & Minutes and longer & Content update\\
			\hline
			Satellite orbit & Position changing/task arriving & N/A & Link scheduling\\
			\hline
			\hline
		\end{tabular}
	\end{table}

The aforementioned SFC-based reconfiguration should be performed at different time-scales, depending on the proper decisions that trigger the reconfiguration, and the delay required to accomplish the reconfiguration, as illustrated in Table \ref{reconfig_table}. After the bi-directional mission offloading is established, the supporting SFC may have to be re-planned when the distribution of the network resources is changed, or other triggering events happen as listed in Table \ref{reconfig_table}. Note that the reconfiguration process for replanning the SFC takes time, from collecting the necessary system information, uploading and downloading the code to space-air nodes and ground nodes respectively. It is vital to carefully tradeoff the benefits from reconfiguring as better matching to the resource distribution, versus the delay and overhead incurred. In some cases, merely optimizing the resource scheduling can guarantee the performance of mission offloading with low overhead, before the reconfiguration is inevitable. In short, the decision of when should perform reconfiguration or resource scheduling, should be addressed as an online decision optimization, under imperfect system information.

\subsection{Security and Privacy}
When seeking help from other networks via mission offloading, the data and code may be exposed to possibly un-trusted entities, which can cause security and privacy threats to users. This calls for mechanisms to perform autointoxication in the mission offloading process, and to ensure data privacy and security. In addition, at the lower layer, jamming and eavesdropping from malicious entities should also be addressed especially for missions offloaded from the space-air to the ground. To this end, SFC should also be planned jointly considering the possible threats from jamming and eavesdropping. More interestingly, voiding secrecy threats can also be an important motivation of mission offloading. For instance, the UAVs can help to relay the information of a vehicle to another RSU in case that the home RSU is jammed \cite{Xiao18}. 

\section{Summary}
In this article, we have elaborated the bi-directional mission offloading framework in SAGIN, which makes full use of the complementary advantages from the space-air networks and ground networks. The overall architecture of agile mission offloading, and the enabling network reconfiguration framework based on NFV and SFC, have been introduced and validated with a case study, which demonstrates the substantial performance gain in reliability and cost reduction. Finally, we have identified several key research issues to further exploit the benefits of bi-directional mission offloading in SAGIN.

\section*{Acknowledgement}
The authors would like to express their thanks for Prof. Jianhua Lu's inspiration on the idea of reverse offloading. 
This work is sponsored in part by the Nature Science Foundation of China (No. 91638204, No.61871254, No. 61801011, No. 61861136003, No. 61571265, No. 61621091), National Key R\&D Program of China 2018YFB0105005, and Hitachi Ltd.

\vspace{-0.2in}

\begin{IEEEbiographynophoto}{Sheng Zhou} [S'06, M'12] (sheng.zhou@tsinghua.edu.cn)
received his B.S. and Ph.D. degrees in Electronic Engineering from Tsinghua University, China, in 2005 and 2011, respectively. He is currently an associate professor of Electronic Engineering Department, Tsinghua University. His research interests include cross-layer design for multiple antenna systems, vehicular networks, mobile edge computing, and green wireless communications.
\end{IEEEbiographynophoto}

\vspace*{-2\baselineskip}

\begin{IEEEbiographynophoto}{Guangchao Wang} (wgc15@mails.tsinghua.edu.cn)
received the B.S. degree in communications engineering from Beijing Jiaotong University, Beijing, China, in 2015. He is currently pursuing the Ph.D. degree in electronic engineering with Tsinghua University, Beijing, China. Her research interests include space-air-ground integrated network reconfiguration and UAV-aided traffic offloading.
\end{IEEEbiographynophoto}

\vspace*{-2\baselineskip}

\begin{IEEEbiographynophoto}{Shan Zhang} [S'13, M'16] (zhangshan18@buaa.edu.cn)
received Ph.D. degree in electronic engineering from Tsinghua University, Beijing, China, in 2016. She is currently an assistant professor in the School of Computer Science and Engineering, Beihang University, Beijing, China. She was a post doctoral fellow in Department of Electronical and Computer Engineering, University of Waterloo, Ontario, Canada, from 2016 to 2017. Her research interests include mobile edge computing, wireless network virtualization and intelligent management. 
\end{IEEEbiographynophoto}

\vspace*{-2\baselineskip}

\begin{IEEEbiographynophoto}{Zhisheng Niu} [M'98, SM'99, F'12] (niuzhs@tsinghua.edu.cn) graduated from Beijing Jiaotong University, China, in 1985, and got his M.E. and D.E. degrees from Toyohashi University of Technology, Japan, in 1989 and 1992, respectively.  During 1992-94, he worked for Fujitsu Laboratories Ltd., Japan, and in 1994 joined with Tsinghua University, Beijing, China, where he is now a professor at the Department of Electronic Engineering. His major research interests include queueing theory, traffic engineering, mobile Internet, radio resource management of wireless networks, and green communication and networks.
\end{IEEEbiographynophoto}

\vspace*{-2\baselineskip}

\begin{IEEEbiographynophoto}{Xuemin (Sherman) Shen} [M'97, SM'02, F'09] (sshen@uwaterloo.ca)
is a University Professor with the Department of Electrical and Computer Engineering, University of Waterloo, Canada. His research focuses on resource management, wireless network security, social networks, and vehicular ad hoc and sensor networks. He is the Vice President on Publications of the IEEE Communications Society. Professor Shen received the James Evans Avant Garde Award from the IEEE Vehicular Technology Society, the Joseph LoCicero Award in 2015 and the Education Award in 2017 from the IEEE Communications Society. He is a registered Professional Engineer of Ontario, Canada, an Engineering Institute of Canada Fellow, a Canadian Academy of Engineering Fellow, a Royal Society of Canada Fellow, and a Distinguished Lecturer of the IEEE Vehicular Technology Society and Communications Society. 
\end{IEEEbiographynophoto}

\end{document}